\documentclass[twocolumn,showpacs,preprintnumbers,amsmath,amssymb]{revtex4}
\usepackage{graphicx}
\usepackage{dcolumn}
\usepackage{bm}
\begin{document}

\title{Superconductivity above the lowest Earth temperature in pressurized sulfur hydride}

\author{ Antonio Bianconi$^{1,2,3}$, Thomas Jarlborg$^{1,4}$  }

\affiliation{
$^1$ 
RICMASS, Rome International Center for Materials Science Superstripes, Via dei Sabelli 119A, 00185 Rome, Italy
\\
$^2$
 Institute of Crystallography, Consiglio Nazionale delle Ricerche, via Salaria, 00015 Monterotondo, Italy
\\
$^3$
INSTM, Consorzio Interuniversitario Nazionale per la Scienza e Tecnologia dei Materiali, Rome Udr, Italy
\\
$^4$ 
DPMC, University of Geneva, 24 Quai Ernest-Ansermet, CH-1211 Geneva 4, Switzerland
}


\begin{abstract}
A recent experiment has shown
a macroscopic quantum coherent condensate at 203 K, about
19 degrees above the coldest temperature recorded on the Earth, 184 K (-89.2 $°C$,  -128.6 $°F$)
in pressurized sulfur hydride. This discovery is relevant  not only in material science and condensed matter 
but also in other fields ranging from quantum computing to quantum physics of living matter.
It has given the start to a gold rush looking for other macroscopic quantum coherent condensates 
in hydrides at the temperature range of living matter $200<T_c<400K$.
We  present here a review of the experimental results and the theoretical works and we discuss the Fermiology of $H_3$S 
focusing on Lifshitz transitions as a function of pressure. 
We discuss the possible role of the $shape$ $resonance$ near a $neck$ $disrupting$ Lifshitz transition, 
in the Bianconi-Perali Valletta (BPV) theory, for rising the critical temperature 
in a multigap superconductor, as the Feshbach resonance rises the critical temperature in Fermionic ultracold gases
\end{abstract}


\maketitle

In March 2014 Mikhail Eremets and his collaborators recorded in their logbook the first evidence 
of high temperature superconductivity in the high pressure metallic phase of  H$_2$S\cite{droz,droz2}
as he reported at the Superstripes 2015 conference \cite{eremets1}. Independently, at the same time, Yinwei Li et al. \cite{li1} 
published the theoretical prediction that the metallic phase of H$_2$S at high pressure should be a stable 
superconductor with high $T_c$. In the fall 2014  the
 work of Duan et al.\cite{duan} predicted high temperature superconductivity in metallic H$_3$S, formed by pressure induced
disproportion of 2($H_2$S)+$H_2$ at around 100 GPa.
\begin{figure}
\includegraphics[height=6cm,width=7.0cm]{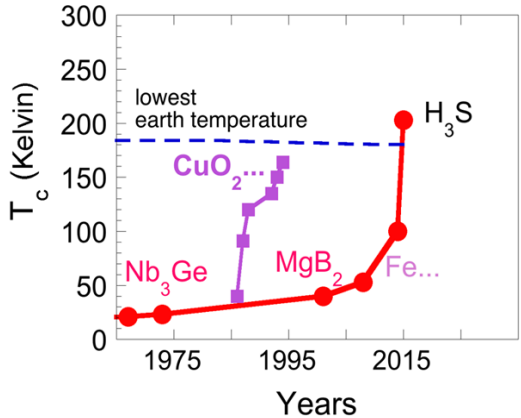}
\caption{(Color online)  Records for the highest superconducting critical temperature found in: cuprates (1994), made of $CuO_2$ layers intercalated by spacer layers, in Mg$B_2$ (2001), in iron
based superconductors (2008), made of $Fe$ layers intercalated by spacer layers,  followed in 2014 by 100K superconductivity in FeSe films on doped SrTi$O_3$.}
\label{Tcyears}
\end{figure}

The record of superconductivity at 190 K in pressurized H$_2$S was reported in December 2014 by evidence of zero resistivity \cite{droz} 
followed by Meissner effect in June 2015\cite{eremets1} with a maximum $T_c$ onset of 203 K  \cite{droz2} establishing a record for
the superconductor with the highest critical temperature. The previous  $T_c$=164K record 
was held by doped HgBa$_2$Ca$_2$Cu$_3$O$_8$, at a pressure of 15 GPa\cite{gao} as 
shown in Fig. 1. The theoretical works on the structure prediction have been performed 
using the USPEX algorithm, Universal Structure Predictor Evolutionary Xtallography, \cite{oga1} already 
applied to predict superconductivity in  LiH$_n$ \cite{oga2}.

\begin{figure}
\includegraphics[height=5cm,width=5.0cm]{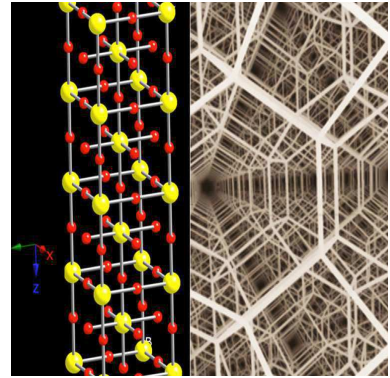}
\caption{(Color online)  
The left panel shows the high pressure $Im\bar{3}m$ phase of H$_3$S. Yellow large spheres indicate sulfur atoms and 
the  red small spheres indicate  hydrogen atoms. The $Im\bar{3}m$ is formed by two intertwinned set of chains made of S-H-S covalent bonds. 
The right panel shows a picture of the complex 3D space filling made of bitruncated cubic tiling at the border line between crystals and quasicrystals.}
\label{structure}
\end{figure}

\begin{figure}
\includegraphics[height=9.0cm,width=6.0cm]{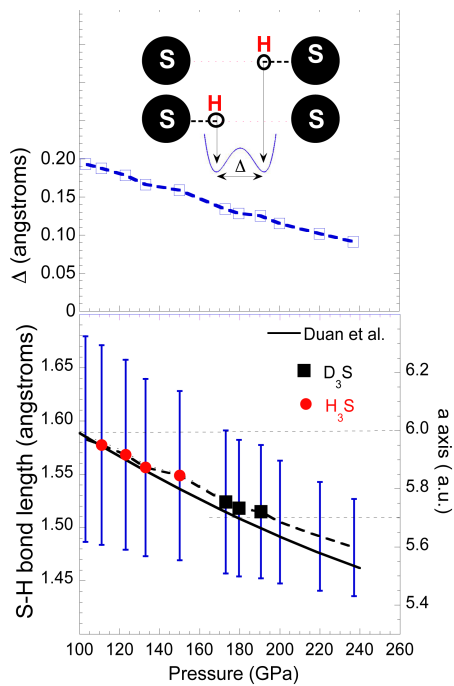}
\caption{(Color online)  
Upper panel: the spatial separation $\Delta$ between the minima of the expected double well potential for hydrogen 
along the S-S direction in $H_3$S typical of the S-H 
hydrogen bond. Lower panel: the S-H bond distances in $H_3$S (filled red dots) and 
$D_3$S (black squares) as a function of pressure\cite{ere0} and 
 the prediction of Duan et al. (black solid line)\cite{duan}.
The error bars show the amplitude of the 
expected fluctuations of the S-H bond calculated by the difference between the two minima of the a 
double well potential for the H atoms.}
\label{bonds}
\end{figure}

\begin{figure}
\includegraphics[height=6.5cm,width=8.0cm]{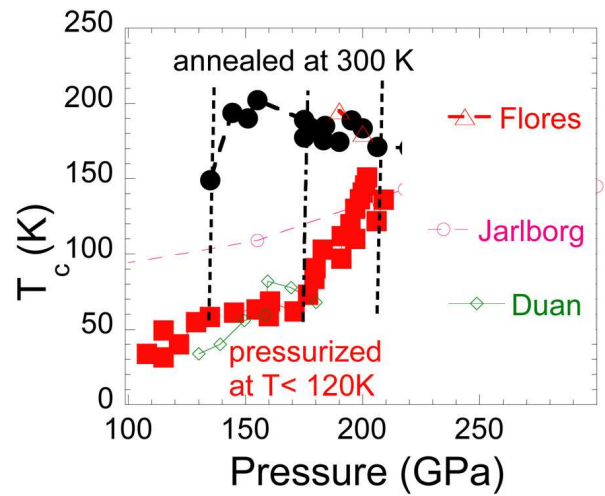}
\caption{(Color online)  
The critical temperature of H$_3$S samples pressurized at low temperature (filled red squares) and annealed at room temperature (filled black dots)
 \cite{droz2,ere0}. The experiment shows a metastable phase in the pressure range between 140 and 205 GPa
where the critical temperature depends on different thermal and pressure treatments.
Calculated critical temperature by Duan et al.\cite{duan}, (open diamonds) Jarlborg et al.\cite{jar}, (open red circles ) Flores et al.\cite{sanna} (open red triangles).}
\label{structure}
\end{figure}

High temperature superconductivity in pressurized $H_3$S has been recently confirmed and the crystal structure has been measured \cite{yLi2,ere0}. 
H$_3$S crystallizes with the $Im\bar{3}m$ lattice symmetry, as predicted \cite{duan}. 
This structure (see Fig.2) is at the border line between crystals and quasicrystals \cite{ste,ram}. In the expected low pressure R3m phase $H_3$S shows
 long and short hydrogen S-H bonds\cite{duan}, like in ice crystals, with a double potential well as shown in Fig. 3.
 The separation $\Delta$ between the minima of the double well decreases 
 with pressure, so the amplitude of H zero point motion decreases with increasing pressure.
The experimental structure\cite{ere0} shows minor divergence from theory predictions\cite{duan} (see Fig. 3).
The system is quite inhomogeneous with broad line-shapes of diffraction profiles becoming sharper at high pressure.  
Moreover the system shows a large increase of the critical temperature by annealing the sample temperature
at 300K, see Fig. \ref{structure}, like in oxygen doped cuprates \cite{poc1}. Therefore an intrinsic optimum inhomogeneity favors superconductivity, like in A15 compounds\cite{a155}, cuprates\cite{campi,poc}, diborides\cite{mgb5} and iron chalcogenides\cite{ricci}.

In the past decades the search for new high temperature superconductors has provided evidence for a large variety of different 
materials with high $T_c$.  A binary intermetallic $Nb_3$Ge, with A15 crystalline structure, has hold the record of T$_c$=23K
for many years\cite{a155}.  This record was exceeded in 1986 by Bednorz and Muller with the discovery\cite{alex} of 35K superconductivity
in doped La$_2$Cu$O_4$, which was followed by the discovery of many superconducting cuprate perovskites
made of CuO$_2$ atomic layers intercalated by a large variety of spacer layers. 
The record of $T_c$=160 K critical temperature in cuprates was achieved by optimization of mainly three physical parameters:\\ 
a) the misfit strain \cite{strain0,strain1,strain2,strain3} between the CuO$_2$ layers and different spacer layers;\\ 
b) the amount of added dopants in the spacer layers (oxygen interstitials) \cite{poc1,poc,campi};\\
c) the application of an external pressure, 15 GPa, in HgBa$_2$Ca$_2$Cu$_3$O$_8$  (Hg1223) \cite{gao}.\\
In 2001 the record for intermetallics was obtained by Akimitsu
and his colleagues measuring transport properties in an intermetallic compound made of light elements Mg$B_2$, 
known since 1953 and already in the market, 
finding  the superconducting transition at $T_c$ =40K \cite{naga}.\\
In 2008 Hosono and his colleagues made the accidental discovery of a new kind of superconducting transition metal oxide, LaAsFeO,
a layered compound doped with fluorine, with $T_c$ = 26 K \cite{kama} which has
triggered the $iron$ age of iron based superconductors
with the record of $T_c$ =100 K in FeSe monoatomic films deposited on doped SrTi$O_3$ \cite{ge}.

Eremets et al. followed for many years  the research direction for high 
temperature superconductivity at high pressure\cite{p1,p2,p3,p4}.
They discovered that metallic silane  becomes superconductor with T$_c$= 17 K at 96  GPa\cite{ere2} and recently they have found that 
P$H_3$  shows  a $T_c$ =100 K at high pressures\cite{ere3}.
Computational materials discovery is now addressed to hydrides considered 
to be pre-compressed phases of solid hydrogen such as
Y$H_4$ and Y$H_6$\cite{Li}, tellurium hydrides \cite{Ma2}, V$H_2$ \cite{Chen}, 
Al$H_3$ \cite{Hou}, Sb$H_4$ \cite{duansb} and polonium hydrides\cite{Liu}.
Superconductivity was predicted in Li$H_6$ with $T_c$ =82 K  at 300 GPa \cite{xie},
in K$H_6$ with $T_c$ =70 K  at 166 GPa \cite{zhou},
with the computed record of $T_c$ =235  in Ca$H_6$ at 150 GPa \cite{wang}.

Many years before high temperature superconductivity in pressurized  
hydrides and hydrogen was proposed by Aschroft \cite{ash1,ash2,ash3,ash4,ash5,Abe},  
Ginzburg  \cite{Ginz1,Ginz2} and Maksimov \cite{mak,mak2}. The theoretical 
predictions of high $T_c$ in solid hydrogen and hydrides at high pressure were 
based on the search of a system with i) a high energy phonon mediating the pairing,
 because of the small mass of the hydrogen ion, with ii) a negative  dielectric 
 constant \cite{Ginz2,mak,mak2} and iii) at the verge of superfluid-superconductor 
 transition \cite{ash5}, where the transition temperature principally increases 
 through the reduction in the associated Coulomb pseudopotential.
 
\begin{figure}
\includegraphics[height=5.71cm,width=7.0cm]{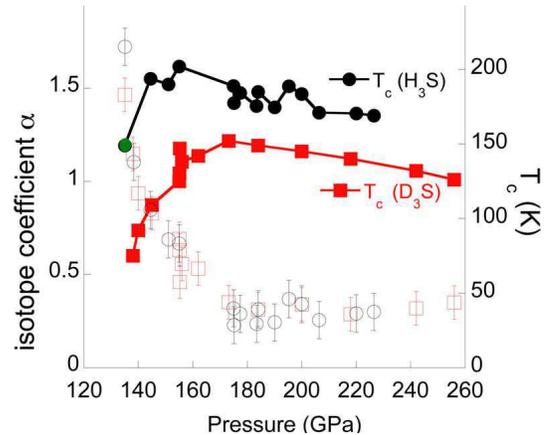}
\caption{(Color online) The isotope coefficient as a function of pressure, (open black circles)
calculated from the critical temperature $T_c$ of H$_3$S (black solid dots) and the interpolated $T_c$ of D$_3$S at the same pressure.
The isotope coefficient, (open red squares)
calculated from the critical temperature $T_c$ of D$_3$S (red solid squares) and the interpolated $T_c$ of H$_3$S at the same pressure  
from the data reported by Drozdov et al. \cite{droz2} and  Einaga et al. \cite{ere0}. }
\label{isotope}
\end{figure}

\begin{figure} 
\includegraphics[height=5.26cm,width=8cm]{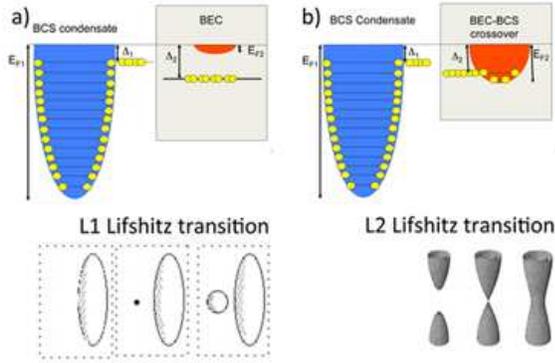}
\caption{(Color online) Pictorial view for the shape resonance described by the BPV theory 
for a two condensates superconductor. The first case where $T_c$ is suppressed is shown in panel a)  with the first condensate in the BCS
 regime and the second condensate in the BEC regime (upper panel) at the Lifshitz transition\cite{bian}, $L1$ type, for the $appearing$ $of$ $a$ $new$ $Fermi$ $surface$ $spot$ (left lower panel).
The second case where $T_c$ is maximum is shown in panel b)  
with the first condensate in the BCS regime and the second condensate 
in the BCS-BEC regime (upper panel) realized at the Lifshitz transition \cite{bian}, $L2$,  for $neck$ $disrupting$ $Fermi$ $surface$ (right lower panel).}
\label{fig_bcsbec}
\end{figure}

The isotope effect in $H_3$S \cite{droz} provided a direct evidence for the conventional phonon mediated pairing. 
Therefore theories of unconventional pairing based on exchange of magnetic interactions have been ruled out. 
 From the results in ref. \cite{droz2}, the isotope coefficient  as a
 function of pressure was reported by \cite{bian}. Fig. \ref{isotope} shows the pressure dependent isotope coefficient from data in ref. \cite{ere0}.
 The isotope coefficient
shows a minimum of $\alpha$=0.2 at 200 GPa, a maximum, $\alpha$=1 at 140 GPa
and a second maximum $\alpha$=0.3 around 240 GPa. A similar pressure dependent isotope
coefficient has been found in cuprates superconductors as a function 
of doping \cite{isotope1,isotope3} with anomalies at Lifshitz transitions  \cite{Lifshitz,Varlamov} of the $L1$ type, $appearing$ $of$ $a$ $new$ $Fermi$ $arc$, or of the $L2$ type, $neck$ $disrupting$ \cite{shape2} (see Fig. \ref{fig_bcsbec}).
Therefore the data in Fig. \ref{isotope} have been interpreted \cite{bian} as indication of the presence of 
Lifshitz  transitions in the pressure range showing high temperature superconductivity.
The  Lifshitz transitions i.e., the electronic transitions in the Fermi surface
 topology of $H_3$S as function of pressure in the range from 80 GPa to 250 GPa have 
 been identified by band structure calculations in ref. \cite{bian,jar}, which have been
 confirmed \cite{qpic}. The giant effect of sample annealing in the range 140-200 GPa in $H_3$S 
 due to the predicted arrested phase separation near Lifshitz transitions \cite{kugel1,kugel2} as for cuprates where intrinsic phase separation 
 and inhomogeneity are key features of high temperature superconductivity \cite{bia1,poc,campi}.
 
The superconducting temperature in sulfur hydride
was interpreted by a series of theoretical papers using the standard BCS approximations.
First, the $dirty$ $limit$ $approximation$ reducing multiple bands, crossing the chemical potential, to a simplified 
metal with a single effective band.
Second, the $Migdal$ $approximation$ considering  the chemical potential very far from band edges so that
the electronic and ionic degrees of freedom can be rigorously separated in agreement with the Born-Oppenheimer approximation.
The superconducting temperature was predicted
by employing the Allen-Dynes modified McMillan formula \cite{duan,duan2},
the Migdal Eliashberg  formula  \cite{papa,errea,dura,nico}, 
and  the more advanced density-functional theory SCDF \cite{sanna,ari}. 
The superconducting condensate has been described in the 
frame of isotropic pairing with a single gap $\Delta_0$ in the BCS regime $\Delta_0/E_F<<1$ and in the Migdal 
approximation $\omega_0/E_F<<1$,  where the Fermi energy $E_F$ is the energy separation between the chemical 
 potential and the bottom of valence bands at -25 eV. 
The pre-factor, $\omega_0$, in the BCS formulas is very 
high of the order of 100-150 meV due to a high energy 
phonon brach formed by hydrogen ions dynamics \cite{errea,sanna,ari}.
However both the electron-phonon coupling and the total density of states (DOS) are not very large.
The phonon energy of 100-150 meV giving the energy cut-off of the pairing interaction in $H_3$S should be compared with the 70 meV 
phonon energy of i) the half-breathing Cu-O-Cu mode, mediating the pairing in cuprates\cite{rez}, and i) the $e_{2g}$ mode in magnesium diboride \cite{sim} 
that show moderate energy softening due to electron-phonon interaction. 
The moderate softening there is due to the fact that these phonons interact only with
small portions of electrons on the Fermi surface: the Fermi arc around the antinodal ($\pi$,0)  point in cuprates, 
and the small tubular $\sigma$ Fermi surface in Mg$B_2$.

Starting from the evidence of Lifshitz transitions the Bianconi, Perali, Valletta (BPV) theory\cite{shape2,shape} 
was proposed \cite{bian,jar} to describe high temperature superconductivity in $H_3$S.
The BPV theory considers high temperature superconductivity made of multiple condensates.
The optimum critical temperature is predicted for 
a first condensate in the BCS regime and 
the second one in the BCS-BEC crossover near a Lifshitz transition of the $neck$ $disrupting$ type as shown in Fig. \ref{fig_bcsbec}.

\begin{figure}
\includegraphics[height=6.0cm,width=7.0cm]{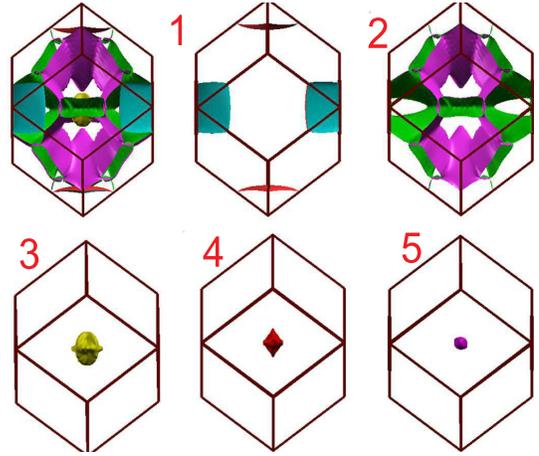}
\caption{(Color online) The Fermi surface for H$_3$S with $Im\bar{3}m$ structure 
for the lattice parameter  $a$= 5.6  a.u. at 200 GPa shown on the top left side of the 
figure. The Fermi surface is formed by 5 different Fermi surfaces from 1 to 5.
The first and second large Fermi surfaces coexist with 3 small Fermi 
surface pockets (No.3,No.4,No.5) centered at the $\Gamma$-point (FS calculations made by Yinwei Li \cite{Li}).}.
\label{figFS}
\end{figure}

The BPV theory is based on the general theory of superconductivity \cite{wei,bla,leg} with no ad-hoc BCS approximations. It was
 proposed in 1996 for cuprates\cite{isotope1}, was verified in 2001 in diborides\cite{mgb7}, and in 2009 for iron based superconductors \cite{iron1,iron3},
where the highest $T_c$ at a $neck$ $disrupting$ Lifshitz transition
 was observed by high resolution ARPES, experiments \cite{iron4,iron5,iron6,iron8}.
Different condensates are formed in different Fermi surfaces with different symmetry and different Fermi energies $En_F$.
The shape resonance provides a key quantum interaction increasing the critical
 temperature, like Feshbach resonance in ultracold gases. 
The shape resonance is a contact interaction not included in the standard Eliashberg theory.
It is an exchange interaction between first pairs in a first BCS condensate and second pairs
in a second condensate in the BCS-BEC crossover regime (see Fig.  \ref{fig_bcsbec}). 
It belongs to the class of Fano-Feshbach resonances widely investigated theoretically and experimentally in ultracold fermionic gases\cite{fes10,fes1,fes2} where
the Feshbach resonance is the only pairing mechanism giving high temperature superconductivity with $k_B$$T_c$/$E_F$=0.2.
Therefore information on the Fermi surfaces are essential for the BPV superconductors made of multiples condensates in multiple Fermi surfaces.
The Fermi surface of H$_3$S at 200 GPa, i.e., for $a$= 5.6  a.u., is shown in Fig. \ref{figFS}. It is formed by
 5 different Fermi surfaces calculated in the bcc Brillouin zone (bcc BZ) like in Y$B_6$  \cite{Li}. 
 
\begin{figure}
\includegraphics[height=9.0cm,width=8.0cm]{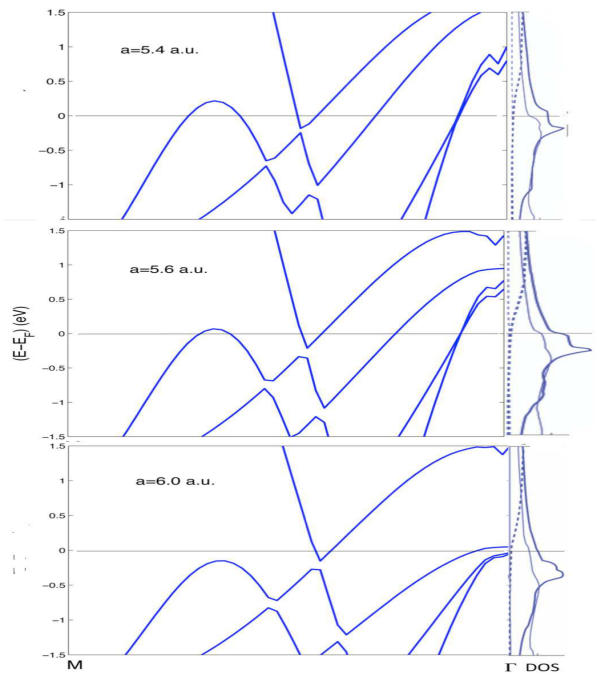}
\caption{(Color online) The band structure along $M-\Gamma$ points of
the simple cubic double cell Brillouin zone (sc BZ) for H$_3$S at different the lattice parameters in the range between 5.4 a.u.
and 6. a.u. The figure shows the shift of the top of a band at about 2/3 in the $M-\Gamma$ direction from low pressure (a=6.0) to high pressure (a=5.4)
and and at 200 GPa, where it is crossing the chemical potential. The top of this band corresponds to a saddle point on the band dispersion.
This crossing point is associated with a neck disrupting Lifshitz transition in the Fermi surface No.2. 
Moreover it is associated 
with the divergent derivative of the partial density of states (DOS) of band No.2
on the high energy side of the sharp DOS peak near the Fermi level.
The partial density of states for each band is shown in the left side of the figure.}
\label{7figXMGR}
\end{figure}

The large Fermi surface(FS) No.2 coexists with the cubic-like FS No.1, and three small closed FS at $\Gamma$ point No.3,4,5.
The band dispersion shown in Fig. 8 has been calculated in the simple cubic Brillouin zone (sc BZ), where the simple cubic unit cell contains 8 sites, 
since it grabs key features of the electronic structure\cite{bian,jar}. The use of sc BZ in Fig. \ref{7figXMGR} is widely used for A15 superconductors. In fact here H 
 ions on the limits of the simple cubic cell forms linear chains, as the transition metals form linear chains in A15 compounds, and  the S sites form a bcc lattice. 
The band dispersion of the second band at about 2/3 of the $\Gamma$-M direction 
shows a maximum above the Fermi level for a=5.6 a.u. as reported in Fig. \ref{7figXMGR}.
The partial density of states only for the bands
crossing the Fermi energy is shown in the panel on the left side of Fig. \ref{7figXMGR}. 
The density of states associated with the No.2 band shows the highest contribution to the total 
density of states (DOS). For $a=5.6$a.u. lattice parameter, we see that the van Hove singularity,
 associated with the derivative peak at the high energy side of the narrow DOS peak, 
 is above the chemical potential, at the same energy position as the top of the band dispersion at 2/3 of the $\Gamma$ -M direction. 
The discontinuity of the DOS indicates a topology change of a portion of the Fermi surface from a 3D topology to 2D topology.
The 2D portions of the Fermi surface appear in the Fermi surface No.2 
in Fig. \ref{figFS} as  tubular necks, in the HNH  direction of the bcc BZ, connecting the large petals of the No.2 Fermi surface.
These tubular necks give the flat energy dependence at the peak of the DOS maximum
near the chemical potential.
The three small Fermi surface pockets (No.3, No.4, No.5) 
centered at the $\Gamma$-point in Fig. \ref{figFS} correspond 
with the three tops at the $\Gamma$-point  of three rapidly dispersing bands  in Fig. \ref{7figXMGR}.
These bands give a small contribution to the total density of states but are 
very sensitive to pressure changes.

\begin{figure}
\includegraphics[height=4.0cm,width=4.0cm]{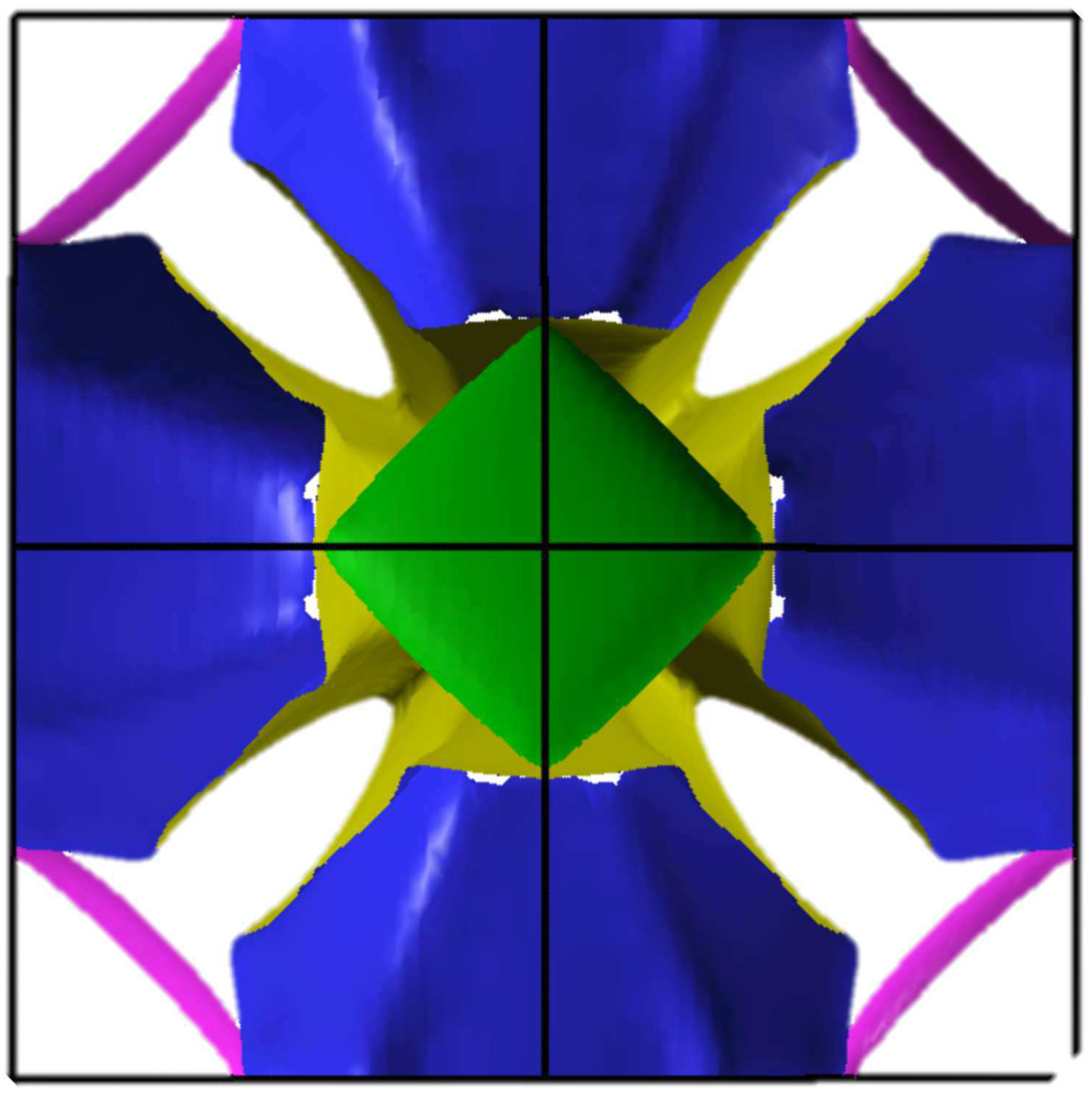}
\includegraphics[height=4.0cm,width=4.0cm]{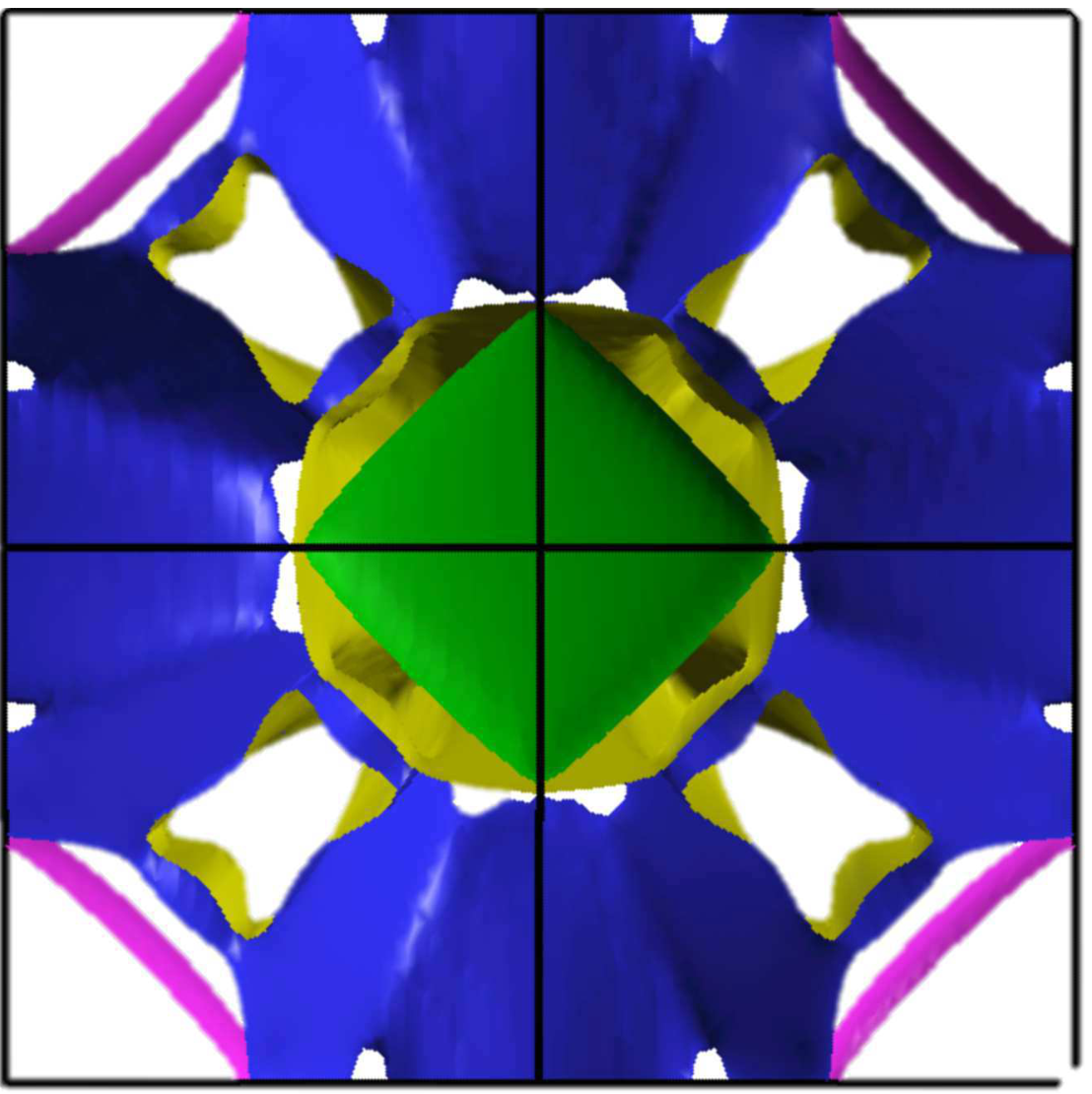}
\includegraphics[height=4.0cm,width=4.0cm]{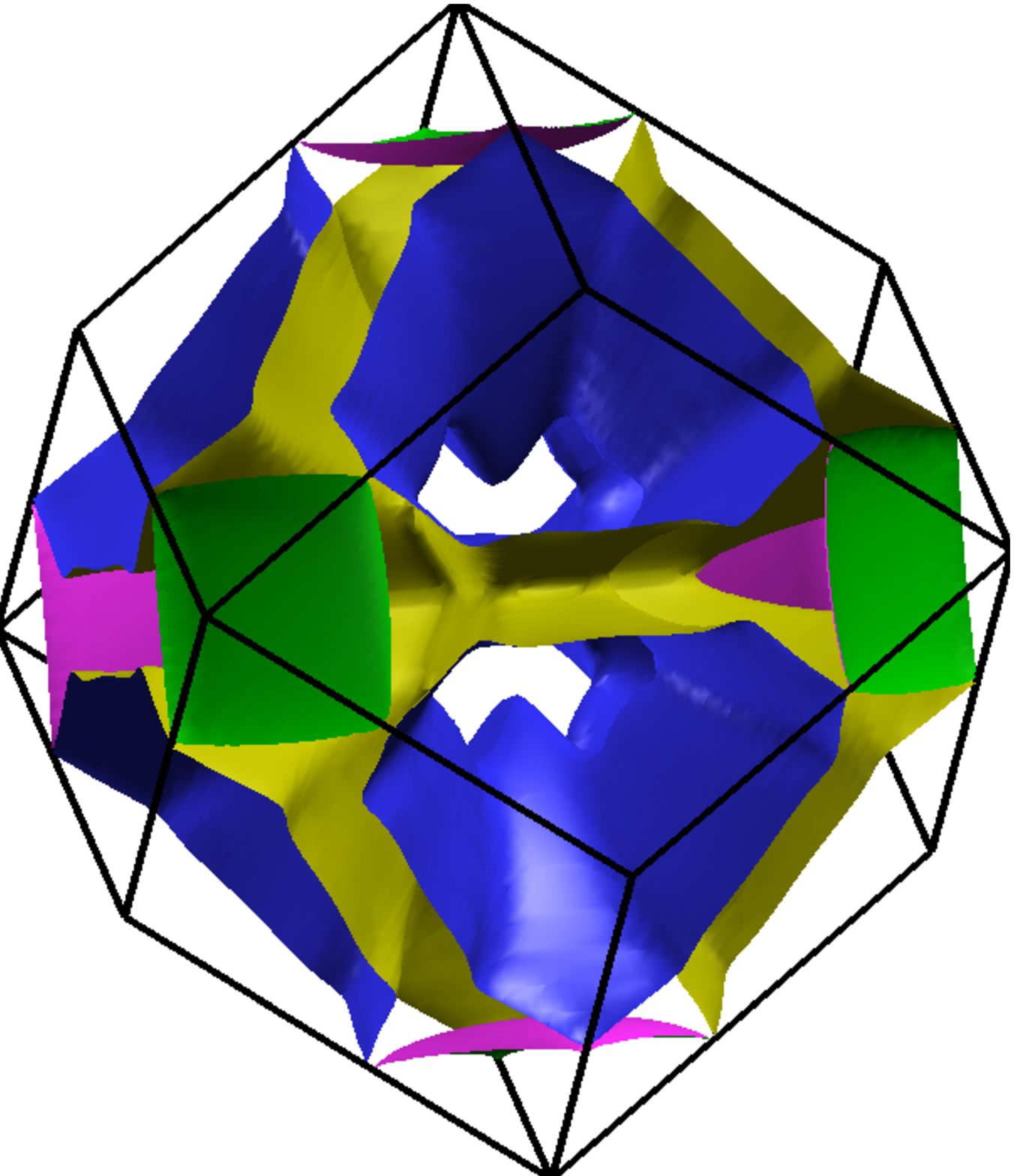}
\includegraphics[height=4.0cm,width=4.0cm]{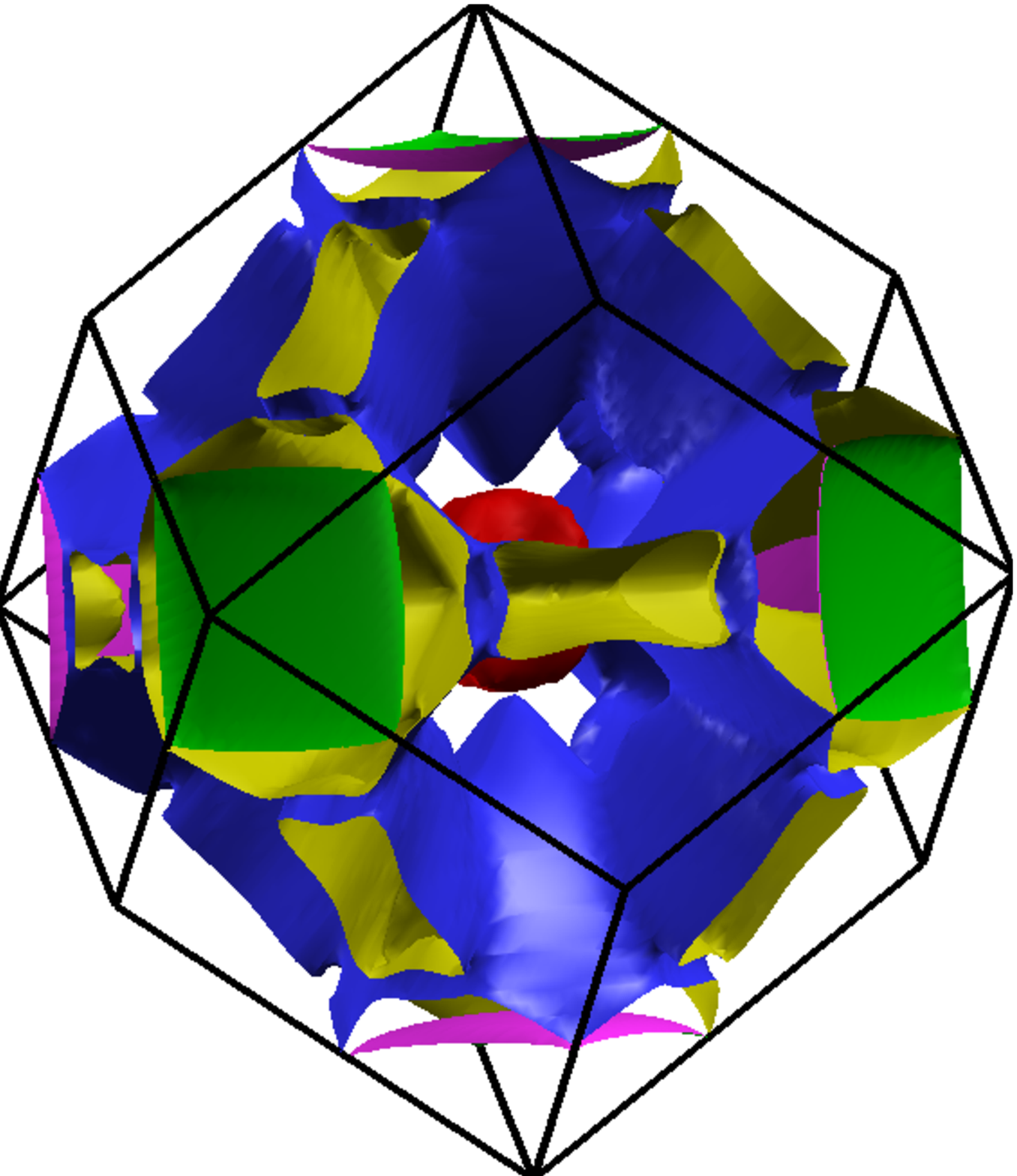}
\caption{(Color online) The two top (bottom) color pictures show the top (side) view of the Fermi surface of H$_3$S 
at low (a=6.0 a.u.) (left panel) and high (a=5.4 a.u.) (right panel) pressure. 
The green and pink surfaces are portions of the Fermi surface No. 1.
The blue and yellow surfaces are portions of the Fermi surface No. 2.
The tubular necks in the Fermi surface connecting 
the large petals of the Fermi surface No.2 appear in the high pressure (a=5.4 a.u.) picture. 
This shows the Fermi surface topology change for a<5.7 a.u. due to the Lifshitz transition for a $Fermi$ $surface$ $neck$ $disrupting, $L2$ type,  $.
The red small Fermi surface pockets No.3,No.4,No.5 centered at the $\Gamma$-point appear for $a<6.0$ a.u. with the change of the Fermi surface topology called Lifshitz transition $L1$ for $new$ $appearing$ $Fermi$ $surface$ $spot$ (FS calculations made by Antonio Sanna \cite{sanna}).}
\label{figsanna}
\end{figure}

\begin{figure}
\includegraphics[height=5cm,width=7.0cm]{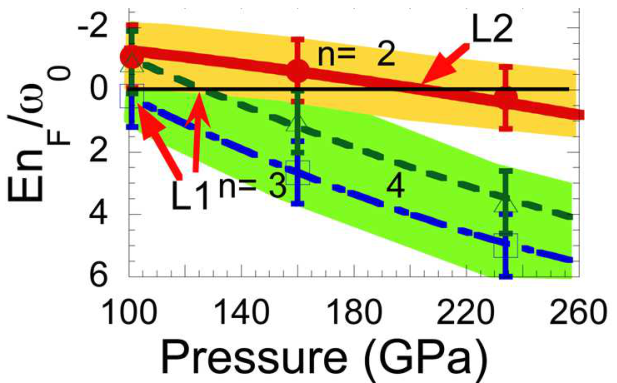}
\caption{(Color online) Blue and green dashed lines show the pressure dependence of
of the Fermi energy for the hole-like bands, No. 3 and No. 4,
crossing the chemical potential at the $L1$ Lifshitz transition, see Fig. \ref{fig_bcsbec}.
The red solid line shows the Fermi energy at the saddle point, at about 2/3 
of the $\Gamma-M$ distance,  crossing the chemical potential at the $L2$ Lifshitz transition, see Fig.  \ref{fig_bcsbec}.
The Fermi energy is divided by the effective pairing energy cut off (taken here to be 150 meV)  due to phonon frequency and zero point motion energy fluctuations \cite{jar} as in diborides \cite{mgb4,mgb5,boeri} 
The errors bars and the colored region indicates the energy fluctuation range of the band edges.}
\label{fig_liftshitz1}
\end{figure}

The Lifshitz transitions driven by pressure are shown in Fig. \ref{7figXMGR} and in Fig. \ref{figsanna}.     
The tops of the dispersing bands at the $\Gamma$-point in
Fig. \ref{7figXMGR} are pushed up by pressure. 

The  $L1$ Lifshitz transitions for a $new$ $appearing$ $Fermi$ $surface$ $spot$
occurs where the bands at the $\Gamma$-point cross the chemical potential.
The side view pictures of all Fermi surfaces in Fig. \ref{figsanna} show 
that the small red closed Fermi surfaces at $\Gamma$ are not present for a=6.0 a.u. while 
they appear at high pressure as a red sphere, as shown for a=5.4 a.u..
The energy shift of the tops of the hole-like bands No. 3, No. 4 at the $\Gamma$
point are shown in Fig. \ref{fig_liftshitz1}.
The $L1$ Lifshitz transitions for the appearing of the new Fermi surface 
spots at $\Gamma$ occurs around 100 GPa, as shown in Fig. \ref{fig_liftshitz1} at the onset of the observed superconducting phase where $T_c$ is close to zero as predicted by the BPV theory .

The $L2$ $Lifshitz$ $transition$ for $neck$ $disrupting$ (see Fig. 6) driven by pressure  is due to the top of 
the band at 2/3 in the $\Gamma$-M direction in Fig. \ref{7figXMGR} moving from
below to above the chemical potential with increasing pressure.
The top view of Fig. \ref{figsanna} shows the formation of the tubular 2D necks missing at a=6.0 a.u. appearing for $a<5.7$ a.u.. 
These necks connect the large Fermi surface petals as seen in the top view of Fig. \ref{figsanna}.
The energy of the top of the band at about 2/3 of the $\Gamma$-M direction corresponds 
with the energy of the discontinuity in the high energy side of the narrow  DOS peak 
near the chemical potential indicating the van Hove singularity.
The appearing of the 2D necks, by increasing pressure, Fig. \ref{fig_liftshitz1}.
occurs in the pressure range, around 180-200 GPa, where the high critical temperature
 is maximum and the isotope coefficient is minimum as predicted by the BPV theory \cite{shape2,shape})

The Migdal approximation breakdown for electrons at the Lifshitz transitions and they enter in the
 BEC or BCS-BEC crossover. While Eliashberg theory breakdown, the BPV theory 
 describes the system by solving both the gap and the density equation and by including the chemical potential shifts in the superconducting phase.
The shape resonance emerges between few pairs in BEC or BCS-BEC 
crossover and all other pairs  in BCS regime. The hydrogen zero point motion gives  \cite{jar} a relevant renormalization of the energy levels like electronic
 correlation and the related energy fluctuations\cite{jar}, as in diborides \cite{mgb5,mgb4,boeri}, extend the energy range of shape resonances.
 
 In conclusion superconducting H$_3$S has a complex crystal structure at the borderline 
 with quasicrystals.  The hydrogen bond plays a key role with large H atomic fluctuations.
 The system shows intrinsic inhomogeneity with segregation 
 of sulfur moreover it shows lattice instability been sensitive to thermal annealing processes Fig. \ref{structure},  like in oxygen doped cuprates
 where $T_c$ is increased by thermal annealing and under illumination \cite{poc1}.
 The high frequency phonon mode is a key prefactor but also other factors like an anomalous dielectric constant and heterogeneity
should cooperate to drive superconductivity to high temperature. 
The zero-point motion of the light  H-atoms induces  strong electronic renormalization \cite{jar}.
H$_3$S  is a multiband superconductor with five different Fermi surfaces with first condensates in the BCS regime coexist with 
second condensates in the BCS-BEC crossover regime (located on small Fermi surface 
spots with small Fermi energy).
The shape resonance near a $L2$ Lifshitz transition, neglected in the Eliashberg theory, 
described by the BPV theory including both i) corrections of the chemical potential due to pairing, and
 ii) the configuration interaction between different condensates is expected to play a key role 
 in the road map for theory driven search of room temperature superconductors.

 \acknowledgments
We thanks Mikhail Eremets, Ryotaro Arita, Emmanuele Cappelluti and L. Ortenzi for discussions, and
Antonio Sanna and Yinwei Li for their Fermi surface color plots. We acknowledge financial support of superstripes institute.

\end{document}